\documentclass[12pt]{article}
\usepackage{amsmath,amssymb,amsfonts,amsthm}
\title{Nonlinear Quantum Optical Springs and Their Nonclassical Properties}

 \author{M.J. Faghihi and M.K. Tavassoly
\\
\footnotesize{Atomic and Molecular Group, Faculty  of Physics,
Yazd University, Yazd, Iran}
\\ \footnotesize{e-mail: mktavassoly@yazd.ac.ir  } }
\begin{document}

 \date{\today}
\maketitle \abstract{
 The original idea of quantum optical spring arises from the requirement of quantization of the frequency
 of oscillations in the Hamiltonian of harmonic oscillator. This purpose is achieved by considering
 a spring whose constant (and so its frequency) depends on the quantum states of another system.
 Recently, it is realized that by the assumption of frequency modulation of $\omega$ to $\omega\sqrt{1+\mu a^\dagger a}$
 the mentioned idea can be established. In the present paper, we generalize the approach of quantum optical spring
 with particular attention to the {\it dependence of frequency to the intensity of radiation field}
 that {\it naturally} observes in the {\it nonlinear coherent states}, from which we arrive at a physical system has been called by us as {\it nonlinear quantum optical spring}. Then, after the introduction of the generalized Hamiltonian of nonlinear quantum optical spring and it's solution, we will investigate the nonclassical properties of the obtained states. Specially, typical collapse and revival
 in the distribution functions and squeezing parameters, as particular quantum features, will be revealed.}




 \newcommand{\norm}[1]{\left\Vert#1\right\Vert}
 \newcommand{\abs}[1]{\left\vert#1\right\vert}
 \newcommand{\set}[1]{\left\{#1\right\}}
 \newcommand{\R}{\mathbb R}
 \newcommand{\I}{\mathbb{I}}
 \newcommand{\C}{\mathbb C}
 \newcommand{\eps}{\varepsilon}
 \newcommand{\To}{\longrightarrow}
 \newcommand{\BX}{\mathbf{B}(X)}
 \newcommand{\HH}{\mathfrak{H}}
 \newcommand{\D}{\mathcal{D}}
 \newcommand{\N}{\mathcal{N}}
 \newcommand{\la}{\lambda}
 \newcommand{\af}{a^{ }_F}
 \newcommand{\afd}{a^\dag_F}
 \newcommand{\afy}{a^{ }_{F^{-1}}}
 \newcommand{\afdy}{a^\dag_{F^{-1}}}
 \newcommand{\fn}{\phi^{ }_n}
 \newcommand{\HD}{\hat{\mathcal{H}}}
 \newcommand{\HDD}{\mathcal{H}}


 \section{Introduction}\label{sec-intro}
 Quantization of harmonic oscillator's Hamiltonian with the definitions
 of creation and annihilation bosnic operators is achieved. The equally-distance spectrum of the system is described by
 $(n+\frac{1}{2})\hbar\omega$.
 An important point is that in quantum and classical Hamiltonians the frequency of
 oscillations ($\omega=\sqrt{k/m}$) is not quantized. The main idea of {\it quantum optical spring} lies in the
 quantization of frequency of oscillations.
 In a sense, one may recognize that this is a
 further quantization which is imposed on the Hamiltonian (or strictly speaking, on the spring constant)
 of the harmonic oscillators.
 Recently, Rai and Agarwal have designed a quantum optical spring
 such that spring constant depends on the quantum state of another system in a special form {\cite{Rai}}.
 In fact, they followed this idea by replacing $\omega$
 with $\omega\sqrt{1+\mu n}$, where $n=a ^ \dag a$ is the number operator.
 Then, the Hamiltonian of their quantum optical spring is introduced (in a special way) as
 \begin{equation}\label{Hmanko}
 H=\frac{p^{2}}{2m}+\frac{1}{2}m\omega^{2}(1+\mu n)x^{2}.
 \end{equation}
 The factor which generalizes spring constant has been called {\it quantized source of modulation} (QSM); which  in this case is $\mu n$.
 The above Hamiltonian suggests that, the eigenstate of the whole system is obtained by
 multiplying the number state with the eigenfunctions of harmonic oscillator {\cite{Rai}}.

 Generally, systems whose frequencies depends on the quantum
 states of another physical system are known.
 The Hamiltonian (\ref{Hmanko}) also shows that quantum optical
 spring is a spring whose frequency depends on the number states of a second
 oscillator (system) in a particular manner.
 So,  the approach of quantum optical spring,
 has  it's root simply in the {\it intensity dependence of
 frequency} of the oscillator.
 But, this is  a quite well-known phenomenon for people who work in quantum optics field \cite{manko1,Matos1996}.
 Briefly speaking, recall that {\it nonlinear coherent states} are natural generalization of canonical
 coherent states (corresponding to linear harmonic oscillator) to generalized coherent
 states (associated with nonlinear oscillators), showing special kind of nonlinearities in vibrational modes.
 Nowadays,  there are very many generalized coherent states which
 can be classified in this important category of quantum states \cite{ali,tavassoly,Roknizadeh,sivakumar}.
 Interestingly, these states also constitute a noticeable part of nonclassical light fields.
 Anyway, nonlinear coherent states are quantum states of a radiation filed whose
 frequencies depends on the intensity of light.
 As one of the most popular examples, we may refer to the $q$-oscillator which was appropriately interpreted as a nonlinear
 oscillator with a particular type of nonlinearity \cite{manko2}.
 In such a system, the frequency of vibrations depends on the energy of the vibrations
 via hyperbolic cosine functions which explicitly depend on the nonlinearity parameter of $q$-oscillator.
 This strange physical feature is elegantly recognized in Ref. \cite{manko2} by the phenomenon has been named {\it blue shift}.

 Based on the above realizations, we motivate to establish a natural
 link between the "quantum optical spring" of Re. [1] and "nonlinear coherent states",
 which lead us to the generalized notion of "{\it nonlinear quantum optical spring}".
 So, the main goal of the present paper lies in the fact that, coupling the
 harmonic oscillator to a QSM (as $\mu n$ presented in \cite{Rai}), may be
 performed in a "natural" manner and at the same time "general" regime to
 nonlinear coherent states associated with nonlinear oscillators.
 There are vast classes of nonlinear coherent states,
 each set characterizes by an operator valued function $f(n)$ (sometimes has been called the deformation
 function). Therefore, unlike the unique case introduced by Rai
 and Agarwal, there are many distinct types of "nonlinear quantum optical
 spring" which are possible to be constructed.
 In addition, it is worth to mention that, due to the previously known relation between
 discrete spectrum of solvable quantum systems with nonlinearity functions of
 nonlinear coherent states, expressed simply by the relation $e_n=nf^2(n)$ \cite{Roknizadeh}, the
 coupling of harmonic oscillator to any quantum system with
 {\it known discrete spectrum} may also be established, appropriately.  So, new
 different types of nonlinear quantum optical springs
 (corresponding to real physical systems such as
 Hydrogen-like spectrum, P\"{o}schl-Teller potential, anharmonic oscillators  and so on) can be
 constructed and discussed, too.
 Thus, our approach to nonlinear quantum optical spring which we suggest here is somewhat different from the
 one suggested in \cite{Rai}. Although, it can recover the case
  of Ref. [1] as a special example.
  Briefly, two remarkable features of our presented work is
  noticeable:
  (i) introducing the "nonlinear quantum optical
  spring" by establishing a natural link between "quantum optical
  spring" and the "nonlinear coherent states", which is a well-known field in quantum optics, and
  (ii) the generality of the proposal and it's potentiality to construct
  various new types of "nonlinear quantum optical  spring", either
  associated with known nonlinearities $f(n)$ or corresponding to discrete spectrum of physical systems
  $e_n$.  This latter advantage is due to the fact that,
  there exists  various classes of known physical systems, that may play the role of secondary
  systems, and couples to the original harmonic oscillator.
  Therefore, some new classes of nonclassical states may be
  generated following our proposed formalism.

 The paper organizes as follows. In the next section,
 after introducing the nonlinear quantum optical spring and  obtaining it's
 Schr\"{o}dinger equation solution, the place of the optical spring of
 Rai and Agarwal in our general approach will be established. In
 section 3 the dynamical behavior of the states of the introduced system will be
 discussed. Quadrature squeezing criteria of entangled states of the
 nonlinear quantum optical spring will be derived in section 4.
 Nextly, in section 5,
 as some applications, the formalism will be applied to a few well-known
 systems and their interesting properties will be studied. Finally, we conclude the paper in section 6.


\section{Introducing nonlinear quantum optical spring}

 The main work of Rai and Agarwal {\cite{Rai}} can be summarized in imposing the {\it intensity dependence} on the {\it frequency} of a
 quantum harmonic oscillator.
 With appropriate physical motivations, the special transformation from $\omega^2$ to $\omega^2(1+\mu n)$ lead to the Hamiltonian
 of quantum optical spring according to relation ({\ref{Hmanko}}), where the authors obtained some elegant and interesting results.
 One can extend the particular transformation mentioned above, to a generalized transformation
 $\omega^2(1+F(n))$, where $F(n)$ is an appropriate function of number
 operator.  Although $F(n)$ may be chosen arbitrarily, but we will
 confine ourselves to some appropriate functions with special physical meaning behind them.
 For this purpose, recall that in the core of the nonlinear coherent states in quantum optics there
 exists an operator-valued function $f(n)$ responsible for the nonlinearity of the oscillator algebras \cite{manko1,Matos1996}.
 These states attracted much attention in recent decades \cite{ali,sivakumar}. As one of the special features of these
 states,
 we may refer to intensity dependence of the frequency of nonclassical lights. This important aspect of
 nonlinear coherent states has been apparently clarified for the $q$-deformed coherent states with a particular
 nonlinearity function {\cite{manko2}}. Therefore, the main goal of our presentation which is to establish a natural link between
 "quantum optical spring" and "nonlinear coherent states" associated to nonlinear oscillator algebra, seems to be possible.
 This idea naturally leads one to a general formalism for {\it "nonlinear quantum optical spring"}.
 It is worth to mention that although we emphasis on the
 nonlinearity of our introduced quantum optical spring, as we will establish
 in the continuation of the paper, the special case introduced by
 Rai and Agarwal is also nonlinear with a special nonlinearity
 function. Accordingly, our work is not a generalization from
 linear to nonlinear quantum optical spring. Instead, along finding
 the natural link between the quantum optical spring and nonlinear
 CSs,  in addition to enriching the physical basis of such systems, a variety of
 quantum optical springs may be constructed.

 Single-mode nonlinear coherent states are known with deformed ladder operators $A=af(n)$ and
 $A ^\dag =f(n)a ^\dag$ where $f(n)$ is an operator-valued function responsible for
 the nonlinearity of the system.
 We assumed $f(n)$ to be real. A suitable description for the dynamics of the nonlinear oscillator is
 \begin{equation}\label{Hdeform}
 H=\omega A ^ \dag A.
 \end{equation}
 Before paying attention to the main goal of the present paper it is reasonable to have a brief
 discussion on the form of the Hamiltonian in the nonlinear
 coherent states approach. The Hamiltonian $H_M=\frac \omega 2 (A ^ \dag A + A A ^ \dag)$ has
 been introduced in \cite{manko1} in analogous to the quantized harmonic
 oscillator formalism. We have previously established in \cite{mancini} that
 requiring the "action identity" criterion on the nonlinear coherent
 states leads us to the simple form of it as expressed in (\ref{Hdeform}).
 This proposal is consistent with the ladder operators formalism and
 Hamiltonian definition have been outlined in {\it supper-symmetric
 quantum mechanics} contexts in the literature \cite{tavassoly,daoud}.
 It is worth also to
 notice that recently a general formalism for the construction of
 coherent state as eigenstate of the annihilation operator of the
 "generalized Heisenberg algebra" (GHA) is introduced
 \cite{Hassouni}. There are some physical examples there. It is easy to
 investigate that the "nonlinear coherent states" for single-mode nonlinear oscillators, as the
 algebraic generalization of standard coherent states, may be consistently placed in GHA structure
 only if one takes the Hamiltonian associated to nonlinear oscillators as in (\ref{Hdeform}). Indeed, one must consider
  $\left\{ A,      A^\dag, J_0\right\}$ with
 $J_0=H$ as the generators of the GHA. After all, our proposal allows
 us to relate simply the nonlinear coherent states to the
 one-dimensional solvable quantum systems with known discrete spectrum, i.e.,
 $f(n)=\sqrt{e_n/n}$, where $H|n\rangle=e_n|n\rangle$.

  Anyway, the time evolution operator
 \begin{equation}
 U(t)=\exp(-iH(n)t/\hbar)
 \end{equation}
 gives the following expression for the time evolved operator $A(t)$
 \begin{equation}
 A(t)=U ^ \dag(t)AU(t)=A\exp(-i\omega\Omega(n)t)
 \end{equation}
 where
 \begin{equation}\label{omega}
 \Omega(n)\equiv(n+1)f^{2}(n+1)-nf^{2}(n).
 \end{equation}
 The latter relation indicates that frequency of oscillations depends explicitly on the intensity in a more general regime.
 Now, we return to the Hamiltonian of harmonic oscillator and deform it as follows
 \begin{equation}\label{myH}
 \HDD=\frac{p^{2}}{2m}+\frac{1}{2}m\omega^{2}\Omega^{2}(n)x^{2}
 \end{equation}
  where the term $\Omega^{2}(n)-1$  in the above Hamiltonian plays the role of QSM.
  The Hamiltonian expression in (\ref{myH}) describes the dynamics of the nonlinear quantum optical spring.
  Before we proceed, it is worth to notice that, while in (\ref{Hmanko}) which originates from \cite{Rai}
   controlling the frequency is limited to number (Fock) states in a special form, by our introduced
   Hamiltonian in (\ref{myH}) this task may be naturally done with either any nonlinear coherent states
   associated with nonlinear oscillator algebra or any generalized coherent states associated with any
   (one dimensional and discrete) solvable quantum system.

 Anyway, using the known solutions
 of harmonic oscillator with eigenfunction $\phi_{n}(x)$  we obtain the solutions of Schr\"{o}dinger equation
 for modulated Hamiltonian as
 $\HDD\psi^{p}_{n}\left|p\right\rangle=E^{(p)}_{n}\psi^{p}_{n}\left|p\right\rangle$
 as
 \begin{eqnarray}\label{shcrodinger}
 \psi^{p}_{n}&=& N_{n}H_{n}(\alpha_{p}x)\sqrt{\frac{\alpha_{p}}{\alpha}}\exp(-\frac{1}{2}\alpha^{2}_{p}x^{2}) \nonumber \\
 E^{p}_{n}&=&\hbar\omega\Omega(p)(n+\frac{1}{2})
 \end{eqnarray}
 where $H_n$ is the $n$th order of Hermite polynomials, $a ^ \dag a\left|p\right\rangle=p\left|p\right\rangle$,
 $\alpha_{p}\equiv(\frac{m\omega\Omega(p)}{\hbar})^{\frac{1}{2}}$, $N_{n}=(\frac{\alpha}{\sqrt{\pi}2^{n}n!})^{\frac{1}{2}}$
 and $\Omega(p)$ introduced in (\ref{omega}).
 Note that $\psi_{n}^{p}$ is the eigenstate of harmonic oscillator with frequency $\omega$ replaced by $\omega\Omega(p)$,
 and the energy eigenvalues of the modulated    Hamiltonian in (\ref{shcrodinger}) characterized  by two quantum numbers $n$ and $p$.
 For a fixed $p$ these states form a complete set. Obviously, from Eq. (\ref{omega})
 it may be seen that if $f(n)=1$ then $\Omega(n)=1$ and so $\psi^{p}_{n}$ simplifies to $\phi_{n}(x)$.
 From now on we will follow nearly similar procedure of Rai and Agarwal with the same initial state for
 the generalized modulated system we introduced
 in (\ref{Hmanko}) as
 \begin{equation}\label{inf}
 \psi(t=0)=\sum_{p,n}C_{pn}\phi_{n}(x)\left|p\right\rangle.
 \end{equation}
 Making use of the time evolution operator with the Hamiltonian in (\ref{myH}) on Eq. (\ref{inf}) we obtain
 \begin{equation}\label{evolution}
 \left|\psi(t)\right\rangle=\sum_{p,n,l}C_{pn}\exp\left(\frac{-iE^{p}_{l} t}
 {\hbar}\right)\left\langle\psi^{p}_{l}|\phi_{n}\right\rangle \left|p\right\rangle \left|\psi^{p}_{l}\right\rangle.
 \end{equation}
 For next purposes it is required to derive the density matrix with the following result
 \begin{eqnarray}\label{density}
 \rho_{0}&=&\sum_{n,l,m,j,p}\left|\psi^{p}_{l}\right\rangle \left\langle\psi^{p}_{j}\right|C_{pn}C ^\ast_{pm}
 \exp\left[\frac{-i(E^{p}_{l}-E^{p}_{j}) t}{\hbar}\right] \nonumber \\
 &\times&\left\langle\psi^{p}_{l}|\phi_{n}\right\rangle \left\langle \phi_{n}
 |\psi^{p}_{j}\right\rangle.
 \end{eqnarray}
 Eq. (\ref{density}) helps us to study the quantum dynamics of the oscillator coupled to QSM. It is easy to check that setting
 \begin{eqnarray}\label{fn-rai}
 f_{RA}(n)&=&\left(\frac {\sum_{j=0}^{n-1}{\sqrt{1+ \mu j}}}{n}\right)^\frac{1}{2} \ \nonumber \\
 &=&\left[\frac{\sqrt{\mu} \left(\zeta(-\frac{1}{2},\frac{1}{\mu})-\zeta(-\frac{1}{2},n+\frac{1}{\mu})\right)}{n}\right]^\frac{1}{2}
 \end{eqnarray}
or equivalently $\Omega_{RA}(n)=\sqrt{1+\mu n}$ in all above relations leads to the recent results of
 Rai and Agarwal in {\cite{Rai}}, where $\zeta(m,n)$ is the well-known Zeta function.
 We would like to end this section with mentioning that choosing different $f(n)$'s leads to distinct nonlinear quantum optical springs.
So our proposal can be actually considered as the generalization of their work.


\section{Quantum dynamics of the nonlinear quantum optical spring}

%
 Let us consider the situation where QSM and the oscillator are respectively prepared
 in a coherent state and in its ground state. Thus, one has
 \begin{equation}\label{inc}
    C_{pn}=\delta_{n0}\frac{\alpha^{p}\exp(-\frac{\left|\alpha\right|^{2}}{2})}{\sqrt{p!}}.
 \end{equation}
 It must be noted the equations (\ref{inc}) and (\ref{inf})
 determine the initial states. So, no relation between these terms
 and the evolution Hamiltionian described the nonlinear oscillator
 may be expected.
 Inserting (\ref{inc}) into Eq. (\ref{density}) we get
 \begin{eqnarray}\label{fdensity}
 \rho_{0}&=&\sum\frac{\left|\alpha\right|^{2p}\exp(-\left|\alpha\right|^{2})}{p!}\exp[-i\omega\Omega(p)t(l-j)]
 \left|\psi^{p}_{l}\right\rangle \left\langle\psi^{p}_{j}\right| \nonumber \\
 &\times& \left\langle\psi^{p}_{l}|\phi_{0}\right\rangle \left\langle
 \phi_{0}|\psi^{p}_{j}\right\rangle.
 \end{eqnarray}
 The probability of finding the oscillator in the initial state is obtained by
 \begin{equation}\label{p0}
 P_{0}(t)=\left\langle \phi_{0}\left|\rho_{0}\right|\phi_{0}\right\rangle=\sum_{p}\left|A_{p}\right|^{2}Q(p)
 \end{equation}
 where $Q(p)=\frac{\left|\alpha\right|^{2p}\exp(-\left|\alpha\right|^{2})}{p!}$ is the Poissonian distribution function and
 \begin{equation}\label{ap}
 A_{p}=\sum_{l}\exp[-i\omega\Omega(p)tl]\left| \left\langle \psi^{p}_{l}|\phi_{0}\right\rangle\right|^{2}.
 \end{equation}
 The latter formula is one of our key results for the description of nonclassical properties of the nonlinear quantum optical spring.
  Now, by calculating $\left\langle \psi^{p}_{l}|\phi_{0}\right\rangle$  in (\ref{ap}) one finally arrives at
 \begin{equation}
 A_{p}=\frac{\left|\beta_{p}\right|^{2}}{\Omega(p)^{\frac{1}{2}}
 [1-(\beta^{2}_{p}-1)^{2}\exp{(-2i\omega\Omega(p)t)}]^{\frac{1}{2}}}
 \end{equation}
 where we set
 \begin{equation}
 \beta^{2}_{p}\equiv\frac{2\;\Omega(p)}{1+\Omega(p)}.
 \end{equation}
 For classical source of modulation it is enough to replace $p$ by $\left|\alpha\right|^{2}$ in Eq.
 (\ref{p0}). Consequently
 \begin{equation}\label{pcl}
 P_{cl}(t)=\frac{\left|\beta_{\alpha}\right|^{4}}{\sqrt{\Omega^{2}(\left|\alpha\right|^{2})(1-2(\beta^{2}_
 {\alpha}-1)^{2}\cos(2\omega_{\alpha}t)+(\beta^{2}_{\alpha}-1)^{4})}}
 \end{equation}
 where $\omega_{\alpha}\equiv\omega\Omega(\left|\alpha\right|^{2})$. Clearly $P_{cl}$ oscillates at frequency $2\omega_{\alpha}$.


\section{ Quadrature squeezing of the states of nonlinear quantum optical spring}

%
%
 Since the number of photons and so the Hamiltonian operator in (\ref{Hdeform}) is constant, we have the following relations
 \begin{eqnarray}\label{xp}
 x(t)&=&x(0)\cos(\omega\Omega(n)t)+\frac{p(0)}{m\omega\Omega(n)}\sin(\omega\Omega(n)t) \nonumber \\
 p(t)&=&p(0)\cos(\omega\Omega(n)t)-m\omega\Omega(n)x(0)\sin(\omega\Omega(n)t).
 \end{eqnarray}
 Now we define the squeezing parameters  as
 \begin{eqnarray}\label{sparameter}
 S_{x}(t)&=& \frac{\left\langle x^{2}(t)\right\rangle-\left\langle x(t)\right\rangle^{2}}{\left\langle x^{2}(0)
 \right\rangle}, \;\;\;\;\;\;\nonumber\\
  S_{p}(t)&=&\frac{\left\langle p^{2}(t)\right\rangle-\left\langle p(t)\right\rangle^{2}}
 {\left\langle p^{2}(0)\right\rangle}.
 \end{eqnarray}
 Paying attention to (\ref{xp}) the latter parameters can be rewritten as follows:
\begin{eqnarray}\label{nsparameter}
  S_{x}(t)&=&\frac{\langle x^{2}(0) \cos^{2}(\omega \Omega(n) t)\rangle + \langle
  (\frac{p(0)}{m \omega \Omega(n)})^{2} \sin^{2}(\omega \Omega(n) t)\rangle}{\langle x^{2}(0)\rangle}, \nonumber \\
  S_{p}(t)&=& ({\langle p^{2}(0) \cos^{2}(\omega \Omega(n) t)\rangle + \langle
  (m \omega \Omega(n)x(0))^{2} \sin^{2}(\omega \Omega(n) t)\rangle}) \nonumber\\&\times& \frac{1}{\langle p^{2}(0)\rangle}.
 \end{eqnarray}
 The expectation values must be calculated with respect to the states in (\ref{inf}). As a result it is easy to show
 \begin{eqnarray}\label{squeezing}
 S_{x}(t)&=&1-\sum_{n}\frac{\left|\alpha\right|^{2n}\exp(-\left|\alpha\right|^{2})\sin^{2}(\omega \Omega({n})t)}{n!}
 \left( 1-  \frac{1}{\Omega^{2}({n})}\right) \nonumber \\
 S_{p}(t)&=&1+\sum_{n}\frac{\left|\alpha\right|^{2n}\exp(-\left|\alpha\right|^{2})\sin^{2}(\omega \Omega({n})t)}{n!} \;
 \nonumber\\ &\times& \left(\Omega^{2}(n)-1\right).
 \end{eqnarray}
 Note that, unlike the special case considered by Rai and Agarwal with $\Omega^2(n)=1+\mu n$ for which $S_{x}$ is always less
 than one ($x$-quadrature is always squeezed) and hence $p$-quadrature is not squeezed at
 all, according to our results, this can not be always so.
 All we may conclude from the two latter relations is that squeezing in both quadratures may be occurred
  (certainly not simultaneously) depending on the selected nonlinearity function $f(n)$.


 \section{Physical applications of the formalism}\label{examples}

 Now, we like to apply the presented formalism to a few classes of nonlinear coherent states
 where the nonlinear optical spring associated with those systems will be clearly obtained. Indeed, there exists
 various nonlinearity functions in the literature which have  been introduced for different purposes,
 mainly due to their nonclassical properties. Among them we only deal with two
 classes of them which seems to be more familiar, i.e., "$q$-deformed coherent states" and
 "center of mass motion of trapped ion".
 \begin{itemize}
 \item{{\it $q$-deformed coherent states}}\\
 As the first example we use the $q$-deformed nonlinearity function. Recall that Man'ko {\it et al} showed
 that $q$-coherent states are indeed nonlinear coherent states with nonlinearity function \cite{manko2}
 \begin{equation}\label{fq}
 f_{q}(p)=\sqrt{\frac{q^{p}-q^{-p}}{p(q-q^{-1})}}=\sqrt{\frac{\sinh(\lambda p)}{p\sinh\lambda}}
 \end{equation}
 where we have set  $q=\ln\lambda$. Clearly, the nonlinearity depends on
 the magnitude of $\lambda$, too. The corresponding states have frequency dependence on the intensity of radiation field (blue shift) \cite{manko2}.
 By replacing Eq. (\ref{fq}) into  (\ref{omega}) one readily gets
 \begin{equation}\label{omegafq}
 \Omega_{q}(p)=\frac{\cosh\left[\frac{\lambda}{2}(2p+1)\right]}{\cosh\frac{\lambda}{2}}
 \end{equation}

 \item{{\it Center of mass motion of trapped ion}}\\
 As the second example for the construction of nonlinear quantum optical spring, we consider the nonlinearity function that is used
 for the description of the motion of a trapped ion \cite{Matos1996}:
 \begin{equation}\label{fti}
    f_{TI}(n, \eta)=L_{n}^{1}(\eta ^{2})\left[(n+1)L_{n}^{0}(\eta ^{2})\right]^{-1}
\end{equation}
 where $L_{n}^{m}$ are generalized Laguerre polynomials of order $m$ and $\eta$ is known
 as the Lamb-Dicke parameter. Note that the nonlinearity also depends on
 the magnitude of $\eta$. By replacing Eq. (\ref{fti}) into  (\ref{omega}) one can readily deduce the
  explicit form of $\Omega_{TI}(p)$ (we do not show it here).
   \end{itemize}
 Our numerical results have been displayed in figures 1 and 2 show respectively for the probability
 distribution function and squeezing parameter in $x$-quadrature  against $\tau$ for $q$-deformed
  coherent states with $\Omega_{q}(p)$ introduced  in (\ref{omegafq}).  Notice that we have used
  some fixed parameters  in all figures (also notice that the mean photon number is indeed $|\alpha|^2$
  and $\tau =  \omega t/(2\pi)$)). A typical collapse and revival exhibition may be observed from the two
  figures,  which is due to the discrete nature of the quantum state of the source of modulation (quantization
  of the frequency of the oscillation).
 In figure 3 (for center of mass motion of trapped ion) we displayed the probability
 distribution as a function of $\tau$ for the different chosen parameters.
  In figure 4 (for the same system) the squeezing parameter in $x$-quadrature have been shown against $\tau$.
  Collapse and revival exhibition can be observed from figures of the latter system, too, which again have it's
  roots in the fact that the spring constant is controlled by another quantum source (QSM).
  To this end we would like to mention that, as we stated before
  all cases considered by us and the special case of Rai and Agarwal
  quantum optical springs are "nonlinear" in nature according to
  our terminology, so the general features of all are the same, at least
  qualitatively.

\section{Conclusion}\label{examples}

   The presented formalism in the present manuscript compare to Rai and Agarwal  method in Ref. [1] has the advantage that the quantization
   of the frequency of oscillations which comes out naturally from the nonlinear coherent states approach were successfully imposed
   on the quantized Hamiltonian of harmonic oscillator.
   So, in view of our presented approach, one has the possibility of imposing a further quantization on the
   quantized harmonic oscillator in very many ways.
   In fact, our formalism can be easily used for a wide range of nonlinear
   oscillators as well as every solvable quantum systems, due to the simple relation $e_{n}=nf^{2}(n)$
   \cite{tavassoly,Roknizadeh,our}. This proposal leads to squeezing in one of the quadratures of the field, as
   well as typical collapses and revivals in probability
   distribution function and squeezing parameter for the states of the "nonlinear quantum optical spring". The latter results indicate
   that the states corresponding to nonlinear quantum optical spring are purely quantum mechanical states, i.e., they are nonclassical states.
   This is consistence with the results of [1], as one may expect.
   So, all we have done is that,
   through paying attention to the basic motivation of the introduction of "quantum  optical spring", i.e.,
   quantization of frequency in a particular manner,
   we found it's most "general" and "fundamental" place in the "nonlinear coherent states" approach in quantum optics. In the light of
   setting this link, we have introduced the general notion of "nonlinear quantum optical spring",
   through which a variety of new and novelty nonclassical states may be
   obtained. The increased interest of physicists in the field of quantum optics for generating the quantum states
   possessing nonclassicality features is another demonstration for the origin of our presentation.
 \begin{flushleft}
 {\bf Acknowledgements}\\
 \end{flushleft}
 One of us (M.J.F.) would like to acknowledge useful discussions with Dr G.R.
 Honarasa and O. Safaeian.


\newpage

{\bf FIGURE CAPTIONS:}

Fig1. caption: The variation of $P_{0}$ as a function of $\tau$
with parameters $\lambda=0.15$
 and $\alpha=2$ for QSM of $q$-deformed coherent states.

Fig2. caption: {The variation of $S_{x}(t)$ as a function of
$\tau$ with parameters $\lambda=0.15$ and $\alpha=2$
 for QSM of $q$-deformed coherent states.}

Fig3. caption: { The variation of $P_{0}$ as a function of $\tau$
with $\eta=0.2$ and $\alpha=1.25$ for QSM of center of mass motion
of trapped ion. }

Fig4. caption: { The variation of $S_{x}(t)$ as a function of
$\tau$ with
 $\eta=0.2$ and $\alpha=1.25$ for QSM of center of mass
motion of trapped ion. }

  \end{document}